\documentstyle[12pt]{article}

\title{\bf Phase diagram and optical conductivity of the one-dimensional
           spinless Holstein model}
\author{Q. Wang{\footnotemark[1]} {} {} and {} H. Zheng\\
     \normalsize \em
     Department of Applied Physics, Shanghai Jiao Tong University,
     Shanghai 200030, P.R.China\\ 
     \normalsize \em 
     Physics Department, Xinjiang Normal University,
     Uromuq 830053, P.R.China\\
         M.Avignon\\
     \normalsize \em
     Laboratoire d'Etudes des Proprietes Electroniques des Solides, \\
     \normalsize \em
     Centre National de la Recherche Scientifique, B.P.166, 
     38042 Grenoble Cedex 9, France}
\footnotetext[1]{E-mail: hzheng@online.sh.cn}
\date{}
\begin{document}
\maketitle
\begin{abstract}
The effects of quantum lattice fluctuations on the Peierls transition and
the optical conductivity in the one-dimensional Holstein model of spinless
fermions have been studied by developing an analytical approach, based on
the unitary transformation method. We show that when the
electron-phonon coupling constant decreases to a
finite critical value the Peierls dimerization is destroyed by the
quantum lattice fluctuations. The dimerization gap is much more reduced by
the quantum lattice fluctuations than the phonon order parameter. The
calculated optical conductivity does not have the inverse-square-root
singularity but have a peak above the gap edge and there exists a
significant tail below the peak. The peak of optical-conductivity spectrum
is not directly corresponding to the dimerized gap. Our results of the
phase diagram and the spectral-weight function agree with those of the 
density matrix renormalization group and the exact diagonalization methods.
\\
\newline
PACS number(s): 71.38.+i; 64.60.Cn; 71.30.+h; 71.45.Lr; 71.35.Cc\\
\end{abstract}
\newpage

  A great deal of quasi-one-dimensional materials, for example, the
halogen-bridged mixed-valence transition-metal complexes, the
conducting polymers, the organic and inorganic spin-Peierls systems, exhibit
an instability against a periodic lattice distortion due to the Peierls
dynamics. Among the models for one-dimensional systems the Holstein
Hamiltonian\cite{refa1} is a typical electron-phonon
coupling model studied by many previous authors. An interesting and
still controversial problem is how the dimerized ground state is modified
when quantum lattice fluctuations are taken into account. The
quantum lattice fluctuations could have an important effect in most
quasi-one-dimensional materials with a dimerized ground state because the
lattice zero-point motion is often comparable to the amplitude of the Peierls
distortion\cite{refa2}. The challenge of
understanding the physics of quantum lattice fluctuations
led to an intense study of the Holstein model.
Generally speaking, the nonadiabatic effect  
suppresses the order parameters of the system\cite{refa3}. As the
optical-absorption concerned, the results of adiabatic
approximations have inverse-square-root singularity at the gap edge. 
However, this approach is questionable and it has been shown that
the quantum lattice fluctuations must be taken into account
to satisfactorily describe some physical properties of
quasi-one-dimensional systems\cite{refa9}.
By considering the nonadiabatic effect, the singularity
may disappear\cite{refa4}. The influences of the phonon frequency on the
optical-conductivity spectrum in the range from $\omega_{0}=0$ to
$\omega_{0}\rightarrow\infty$  should be
studied for understanding the physics of electron-phonon 
interactions in nonadiabatic case.

When the quantum lattice fluctuations are taken into account
the theoretical analysis
becomes much more difficult. In the past several years, the Holstein 
Hamiltonian has been investigated by using various numerical approaches,
such as Green's function Monte Carlo simulation\cite{refb1,refb2},
renormalization group analysis\cite{refb3,refb4}, variational method of
the squeezed-polaron wave-function\cite{refb5}, phenomenological
approach\cite{refa2}, exact diagonalization\cite{refb7}, etc.
Very recently, works of numerical approaches have been performed in relation
to the Peierls transition and the optical conductivity in the
one-dimensional Holstein model of spinless
fermions by using the density matrix renormalization group\cite{refa5} and
the exact diagonalization\cite{refa6} methods. However, as was pointed
out in Ref.\cite{refa6}, because of the effects of limited system sizes
in numerical approaches, the precise determination of the
critical value in the small $\omega_{0}$ regime is somewhat difficult and
the precise extraction of the dimerized gap from optical
conductivity data is prevented. An analytical study of
the Holstein model will make it possible to have an
insight into the intrinsical properties of the molecular crystal materials.
In a resent work, two of us\cite{refa7} studied this model and investigated
the dimerization order parameters and density
of states in gapped phase and the velocity of
charge excitations and Luttinger liquid stiffness constant
in gapless phase.
In this work we concentrate on the properties of the phase transition and
the optical-responses of the system with the view of understanding the
effects of quantum lattice fluctuations on the Peierls
instability and the optical conductivity in the Holstein
model. We will show that our results of the
phase diagram and the spectral-weight function agree
surprisingly well with those of the density matrix renormalization
group\cite{refa5} and the exact diagonalization\cite{refa6} methods, 
and in our theory the critical value can
be determined precisely even for extremely small phonon frequency.
The effects of quantum lattice fluctuations on the dimerized gap and on the
order parameters are essentially different. The peak of
optical-conductivity spectrum is not directly corresponding to the
dimerization gap.

The one-dimensional spinless Holstein model, in momentum space is
\begin{eqnarray}
 H & = & \sum_{q}\omega_{0}b^{\dag}_{q}b_{q} 
         +\sum_{k}\epsilon_{k}c^{\dag}_{k}c_{k}
          -\frac{1}{\sqrt{N}}\sum_{q,k}g(b_{q}+b^{\dag}_{-q})
         c^{\dag}_{k+q}c_{k},
\end{eqnarray}
where $\epsilon_{k}=-2t\cos k$ is the bare band structure, $t$ the
hopping integral, and $N$ the total number of sites.    
$c_{k}$ and $b_{q}$ are the annihilation operators of electrons with
momentum $k$ and phonons with momentum $q$, respectively. The dispersionless
phonon frequency $\omega_{0}=\sqrt{K/M}$ and $g$ is the electron-phonon
coupling, $K$ the elastic constant and $M$
the mass of ions (throughout this paper we set $\hbar=k_{B}=1$).
          
  In order to take into account the
electron-phonon correlation, the unitary transformation approach is used to
treat $H$\cite{refa7}, $\tilde{H}=e^{S}He^{-S}$. After averaging the
transformed Hamiltonian over the phonon vacuum state we get an effective
Hamiltonian for the fermions
\begin{eqnarray}
 H_{\rm eff} & = & \frac{1}{2}KNu_{0}^{2}+\sum_{k}E_{0}(k)c^{\dag}_{k}c_{k}
               -\sum_{k>0}\Delta_{0}(k)(c^{\dag}_{k-\pi}c_{k}
               +c^{\dag}_{k}c_{k-\pi})\nonumber\\
         & & -\frac{1}{N}\sum_{q,k,k'}
               \frac{g^{2}}{\omega_{0}}
               \delta(k+q,k)[2-\delta(k'-q,k')]c^{\dag}_{k+q}c_{k}
               c^{\dag}_{k'-q}c_{k'},
\label{heff}
\end{eqnarray}
where
\begin{eqnarray}
E_{0}(k) & = & \epsilon_{k}-\frac{1}{N}\sum_{k'}
               \frac{g^{2}}{\omega^{2}_{0}}\delta(k',k)\delta(k,k')
                 (\epsilon_{k}-\epsilon_{k'}),\\
\Delta_{0}(k) & = & \alpha u_{0}[1-\delta(k-\pi,k)],
\end{eqnarray}
$\alpha=g\sqrt{2M\omega_{0}}$, and
$\delta(k+q,k)=1/(1+|\epsilon_{k+q}-\epsilon_{k}|/\omega_{0})$ is a function
of the energies of the incoming and outgoing fermions in the electron-phonon 
scattering process.
This effective Hamiltonian works well in the
$\omega_{0}=0$ and $\omega_{0}\rightarrow\infty$ limits.

   $u_{0}$ can be determined by the variational principle,
\begin{equation}
 u_{0}=\frac{\alpha}{KN}\sum_{k>0}[1-\delta(k-\pi,k)]
       \left\langle fe\left|(c^{\dag}_{k-\pi}c_{k}
       +c^{\dag}_{k}c_{k-\pi})\right|fe\right\rangle.
\label{uofun}
\end{equation}
Here $|fe\rangle$ is the ground state of $H_{\rm eff}$. Thus,
the total Hamiltonian can be written as
$\tilde{H}=\tilde{H}_{0}+\tilde{H}_{1}$, where
$\tilde{H}_{1}$ includes the terms which are zero
after being averaged over the phonon vacuum state, and
\begin{equation}
\tilde{H}_{0}=\sum_{q}\omega_{0}b^{\dag}_{q}b_{q}+H_{\rm eff}.
\end{equation} 

By means of the Green's function method to implement the perturbation
treatment on the four-fermion term in
Eq.(\ref{heff}), we get the renormalized band function and the gap
function\cite{refa7}
\begin{eqnarray}
 E_{k} & = & E_{0}(k)-\frac{2t\lambda}{N}\sum_{k'>0}
              \{\delta(k',k)[2-\delta(k',k)]
              -\delta(k'-\pi,k)[2-\delta(k'-\pi,k)]\}\nonumber\\
      & & \times\frac{E_{0}(k')}{\sqrt{E_{0}^{2}(k')+\Delta^{2}_{0}(k')}},\\
\Delta_{k} & = & \alpha u_{0}[c-d\delta(k-\pi,k)].
\label{delta}
\end{eqnarray}
Where, the dimensionless coupling $\lambda=g^{2}/2t\omega_{0}$,
$W_{k}=\sqrt{E^{2}_{k}+\Delta^{2}_{k}}$
is the fermionic spectrum in the gapped state, and 
\begin{eqnarray}
 c & = & 1+\frac{2t\lambda}{N}\sum_{k>0}[\delta(k-\pi,k)-V]
         \frac{\Delta_{0}(k)}{\alpha u_{0}\sqrt{E_{0}^{2}(k)
         +\Delta^{2}_{0}(k)}},\\
 d & = & 1-\frac{2t\lambda}{N}\sum_{k>0}[1-\delta(k-\pi,k)]
         \frac{\Delta_{0}(k)}
         {\alpha u_{0}\sqrt{E_{0}^{2}(k)+\Delta^{2}_{0}(k)}}.
\end{eqnarray}
\begin{eqnarray}
 V & = & \frac{1}{N^{3}}\sum_{q,k,k'}\delta(k,k+q)[2-\delta(k',k'-q)]
\end{eqnarray}
is the on-site interaction which should be subtracted because of the Pauli
principle.

The phonon-staggered ordering parameter is
\begin{eqnarray}
 m_{p} & = & \frac{1}{N}\sum_{l}(-1)^{l}<u_{l}>\nonumber\\
       & = & \frac{\alpha}{KN}\sum_{k>0}\frac{\Delta(k)}{W(k)}.
\label{mppp}
\end{eqnarray}
These are basic equations in our theory.
        
  From Eq.(\ref{uofun}), let $u_{0}=0$, we get the self-consistent equation 
of phase transition points in the $g^{2}/\omega_{0}\sim\omega_{0}$ plane,
\begin{equation}
 1=\frac{4t\lambda}{N}\sum_{k>0}[1-\delta(k-\pi,k)]
         \frac{c-d\delta(k-\pi,k)}{|E_{k}|}.
\label{figu41}
\end{equation}

If $\omega_{0}=0$ we have $\delta(k',k)=0$ and $c=1$, Eq.(\ref{uofun})
becomes the same as that in the adiabatic theory. In our theory
$\delta(k-\pi,k)$ has a sharp peak at the Fermi point and, since
$1-\delta(k-\pi,k)=4t|\cos k|/(\omega_{0}+4t|\cos k|)$, the logarithmic
singularity in the integration of Eq.(\ref{uofun}) in the
adiabatic case is removed as long as the
ratio $\omega_{0}/t_{0}$ is finite. 
Comparing Eq.(\ref{delta}) with that in the adiabatic case, $\Delta=
\alpha u_{0}$, we have the gap in the nonadiabatic case,
\begin{equation}
\Delta=\Delta(\pi/2)=\alpha u_{0}[c-d].
\label{fig2}
\end{equation}
This is the true gap in the fermionic spectrum.

  Fig.1 shows the ground state phase diagram in the
$g/\omega_{0}\sim t/\omega_{0}$ plane. The solid line is our analytical
result. For comparison, the results of previous authors are also shown.
The line with circle and the line with square denote the results of the
density matrix renormalization group (DMRG)\cite{refa5} and the two-cutoff
renormalization group(TCRG)\cite{refa13}, respectively. The
line with triangle is the phase boundary of the variational Lanczos approach
based on the inhomogeneous
modified variational Lang-Firsov transformation (IMVLF)\cite{refa6}.
To check the consistency of the phase transition quantitatively with that of
DMRG, some critical values $g_{c}$ are listed in Table \ref{tab1}. 
One can see from both the figure and the table that our results agree
surprisingly well with that of DMRG except for very large $\omega_{0}$
($\omega_{0}/t\ge 10$).
Furthermore, our theory can get the phase boundary, separating Luttinger
liquid (LL) and insulation (CDW) phases, even in the very small
$\omega_{0}$ regime, which is theoretically and experimentally significant
since, from the view point of experiment, for quite a lot of realistic cases 
the frequency of quantum phonon $\omega_{0}$ is small.
It seems that in the DMRG and the finite-lattice Lanczos
approach, because of the effects of limited system sizes, 
the precise determination of the critical value in the small
$\omega_{0}$ regime is somewhat difficult\cite{refa6}.
The infinite system is never really
gapless within the adiabatic approach, because the gap remains nonzero,
although it becomes very small for weak electron-phonon coupling. On the
contrary, in our theory, the logarithmic singularity
$\int^{\pi/2}_{0}dk/\cos k$ in the integration of
Eq.(\ref{uofun}) is removed by the factor
$1-\delta(k-\pi,k)$ and the critical value $\lambda_{c}$ can
be determined precisely even for extremely small phonon frequency.

Inspired by the success of obtaining the phase diagram, we further
investigate the dimerization gap and the optical responses.

Fig.2 shows the dimerization gap $\Delta/t=\Delta(\pi/2)/t$ and the
phonon order parameter $\alpha  m_{p}/t$ as functions of the phonon frequency
$\omega_{0}/t$ for $\lambda=0.81$. It is the most notable that 
there is a discontinuous drop in
the dimerization gap once the phonon frequency changes, no matter how small
it is, from zero to finite, though at the adiabatic limit the dimerization
gap $\Delta(\pi/2)/t=\alpha  m_{p}/t$. After the drop the dimerization gap
and the phonon order parameter decrease as the phonon frequency increases.
At the critical value $\omega_{c}$, the dimerization gap and the phonon order
parameter go to zero simultaneously and the system becomes gapless, which
indicates that the quantum lattice fluctuations can destroy the dimerized
Peierls state. The dimerization gap is much more
reduced by the quantum lattice fluctuations than the phonon order parameter.
The reason for the different behavior of the dimerization gap and the
order parameter is that the former is the value of Eq.(\ref{fig2})
at the Fermi point $k=\pi/2$, where the quantum
lattice fluctuations have the strongest effect, while the latter is the
integral (see Eq.(\ref{mppp})) over the all
Brillouin zone and the effect of the quantum lattice fluctuations is
gentled. The effects of quantum lattice fluctuations on the dimerization
gap and on the phonon order parameter are essentially different, especially
when $\omega_{0}$ is small.
In the mean field (MF) 
approximation the Peierls distortion opens a gap $2\Delta_{\rm MF}$ and
$\Delta_{\rm MF}=\alpha m_{p}$. This relation is sometimes assumed remains
valid when quantum lattice fluctuations are taken into account.
Our results indicate that this relation holds only in the
adiabatic limit.

The optical conductivity $\sigma(\omega)$ can be expressed by the
retarded Green's function as follows:
\begin{equation}
\sigma(\omega)=-\frac{2\varepsilon_{0}nc}{\pi\omega}{\rm Im}K^{R}(\omega),
\end{equation}
where $K^{R}$ is defined as
\begin{equation}
 K^{R}(\omega)=-i\int^{0}_{-\infty}e^{-i\omega t}dt
                   \langle g|[j(0)j(t)-j(t)j(0)]|g\rangle.
\end{equation}
Here $j$ is the current operator\cite{refa14},
\begin{eqnarray}
 j & = & -ieta\sum_{l}(c^{\dag}_{l}c_{l+1}-c^{\dag}_{l+1}c_{l}),
\end{eqnarray}
and $j(t)=\exp(iHt)j\exp(-iHt)$ is the form of $j$ in the
Heisenberg representation. The unitary
transformation of current operator is $e^{S}je^{-S}=j+[S,j]
+{{1}\over{2}}[S,[S,j]]+O(g^{3})$. All terms of higher order than
$g^{2}$ will be omitted in the following treatment.
Because the averaging of $\tilde{H_{1}}$ over the phonon vacuum state is
zero, in the ground state at zero temperature $\tilde{H_{1}}$ can be
neglected. By using the approximately decoupling
$|g'\rangle\approx|g'_{0}\rangle$, the ground state of $\tilde{H_{0}}$,
and $\tilde{H}\approx\tilde{H}_{0}$ in the calculation
\begin{eqnarray}
 \langle g|j(0)j(t)|g\rangle & = & \langle g'|[e^{(S+R)}je^{-(S+R)}]
                       e^{i\tilde{H}t}[e^{(S+R)}je^{-(S+R)}]e^{-i\tilde{H}t}
                       |g'\rangle\nonumber\\
          & \approx & \langle g'_{0}|[e^{S}je^{-S}]e^{i\tilde{H}_{0}t}
                      [e^{S}je^{-S}]e^{-i\tilde{H}_{0}t}|g'_{0}\rangle,
\end{eqnarray}
we can get
\begin{eqnarray}
K^{R}(\omega)&=&\frac{J^{2}}{N}\sum_{k>0}\left(\frac{1}{\omega-2W_{k}+i0^{+}}
                 -\frac{1}{\omega+2W_{k}-i0^{+}}\right)
                 \sin^{2}k\left[1-\frac{2}{N}
                 \sum_{k'}\frac{g^{2}}{\omega_{0}^{2}}\delta^{2}(k',k)\right]
                     \frac{\Delta^{2}_{k}}{W^{2}_{k}}\nonumber\\
               &   & +\frac{J^{2}}{N^{2}}\sum_{k>0,k'>0}
                      \frac{g^{2}}{\omega_{0}^{2}}\left(
                     \frac{1}{\omega-\omega_{0}-W_{k}-W_{k'}+i0^{+}}
                     -\frac{1}{\omega+\omega_{0}+W_{k}+W_{k'}-i0^{+}}
                     \right)\nonumber\\
                & & \times[\delta^{2}(k',k)
                     (\sin k'-\sin k)^{2}(\alpha_{k}\beta_{k'}
                     +\beta_{k}\alpha_{k'})^{2}\nonumber\\
                &  & +\delta^{2}(k'-\pi,k)(\sin k'+\sin k)^{2}
                      (\alpha_{k}\alpha_{k'}+\beta_{k}\beta_{k'})^{2}],
\end{eqnarray}
where $\alpha_{k}=\sqrt{(1+E_{k}/W_{k})/2}$, and
$\beta_{k}=\sqrt{(1-E_{k}/W_{k})/2}$. Thus, the optical conductivity
\begin{eqnarray}
\sigma(\omega)&=&\frac{2\varepsilon_{0}ncJ^{2}}{\omega}\sum_{k>0}
                \delta(\omega-2W_{k})\sin^{2}k\left[1-\frac{2}{N}
                 \sum_{k'}\frac{g^{2}}{\omega_{0}^{2}}\delta^{2}(k',k)\right]
                     \frac{\Delta^{2}_{k}}{W^{2}_{k}}\nonumber\\
               & & +\frac{2\varepsilon_{0}ncJ^{2}}{\omega N}\sum_{k>0,k'>0}
                      \frac{g^{2}}{\omega_{0}^{2}}
                    \delta(\omega-\omega_{0}-W_{k}-W_{k'})[\delta^{2}(k',k)
                     (\sin k'-\sin k)^{2}(\alpha_{k}\beta_{k'}
                     +\beta_{k}\alpha_{k'})^{2}\nonumber\\
                &  & +\delta^{2}(k'-\pi,k)(\sin k'+\sin k)^{2}
                      (\alpha_{k}\alpha_{k'}+\beta_{k}\beta_{k'})^{2}],
\end{eqnarray}
and the $\omega-$integrated spectral-weight function
\begin{eqnarray}
 S(\omega) & = &\int^{\omega}_{0}\sigma(\omega')d\omega'.
\end{eqnarray}

The optical conductivity for different phonon frequencies are shown in
Fig.3. The parameter values
used are: $\lambda=1.0$, with       
$\omega_{0}/t=0.01, 0.05$, and $0.10$. One can see that as $\omega_{0}$
increases, the optical-absorption spectrum broadens but the peak height
decreases and moves to lower photon energy, and the spectral-weight increases
as $\omega_{0}$ increases. The inverse-square-root singularity
at the gap edge in the adiabatic case\cite{refa15} disappears
and there is a significant tail below the peak because of the
nonadiabatic effect. We note that in our theory,
in mathematical viewpoint, the difference between the $\omega_{0}=0$ and
$\omega_{0}>0$ cases mainly comes from the functional form of the gap
[see Eq.(\ref{delta})]. Comparing it with that in the adiabatic limit,
one can see that the subgap states come from the quantum lattice
fluctuations, i.e., the second term in the square bracket of Eq.(\ref{delta}).

The rescaled
$\omega$-integrated spectral-weight function $S(\omega)/S_{\rm m}$
versus the photon energy $\omega/t$ relations of our result (solid line)
and that of IMVLF (dashed line)  are shown in Fig.4, where
$S_{\rm m}=S(\omega\rightarrow\infty)$.
The optical conductivity $\sigma(\omega)$ (rescaled by $S_{\rm m}$) of our
result (solid line) is also shown. Because of the finite-size effects
the optical conductivity of IMVLF is
oscillatory and can not be compared with our result directly. 
We use the same parameter values as in Ref.\cite{refa6} (Fig.6(b)):
$g/\omega_{0}=4.47$, and $\omega_{0}/t=0.1$. The true gap
$2\Delta(\pi/2)/t=1.04$ obtained from Eq.(\ref{fig2}) is marked by the
arrow. One can see that the peak of optical-conductivity spectrum is not
directly corresponding to the dimerized gap. The energy gap is small than
the activation energy of the optical-conductivity.
Our result shows clearly both the position and the peak of
optical conductivity and the
spectral-weight agrees with that of IMVLF. While in the exact
diagonalization method the finite-size effects prevent a precise extraction
of the CDW gap from the optical exact diagonalization data\cite{refa6}.

 In conclusion, the effects of quantum lattice fluctuations on the
optical-conductivity spectrum and the ground state phase diagram
of the one-dimensional Holstein model of spinless
fermions have been studied by developing an analytical approach.
We show that when the electron-phonon coupling constant decreases the
dimerization gap decreases, and at a
finite critical value the Peierls dimerization is destroyed by the
quantum lattice fluctuations. The critical value of
electron-phonon coupling can be determined precisely even for very small
phonon frequency. The dimerization gap is much more
reduced by the quantum lattice fluctuations than the order parameter.
The calculated optical conductivity does not have the inverse-square-root
singularity but have a peak above the gap edge and there exists a
significant tail below the peak. In nonadiabatic case the dimerization gap
is small than what the peak position of the optical
conductivity is corresponding to. Our results of the
phase diagram and the spectral-weight function agree with those of the 
density matrix renormalization group and the exact diagonalization methods.

\newpage
{\bf\Large Figure Caption}\\

{\bf Fig.1} The ground state phase diagram in the
$g/\omega_{0}\sim t/\omega_0$ plane. The solid line is our analytical
result. The line with open circle and the line with open square denote
the results of the density
matrix renormalization group (DMRG) and the two-cutoff renormalization
group(TCRG) methods, respectively. The line with open triangle is the phase
boundary of the variational Lanczos approach (IMVLF).\\

{\bf Fig.2} The dimerization gap $\Delta/t=\Delta(\pi/2)/t$
and the order parameter $\alpha m_{p}/t$ as
functions of the phonon frequency $\omega_{0}/t$ in the case of
$\lambda=0.81$.\\

{\bf Fig.3} The optical conductivity in the case of $\lambda=1.0$ for
different phonon frequencies $\omega_{0}/t=0.01, 0.05$, and $0.10$.\\
                                                     
{\bf Fig.4} The rescaled optical conductivity
$\sigma(\omega)$ and $\omega$-integrated spectral-weight function
$S(\omega)/S_{\rm m}$ versus the photon energy $\omega/t$
relations of our results and that of IMVLF, where
$S_{\rm m}=S(\omega\rightarrow\infty)$.
The parameter values used are same as in Ref.\cite{refa7} (Fig.6(b)):
$g/\omega_{0}=4.47$, and $\omega_{0}/t=0.1$.
The dashed line is the result of IMVLF and the solid lines are our
results. The arrow marks the true gap $2\Delta(\pi/2)/t=1.04$.\\

\newpage
{\bf\Large Table Caption}\\
\begin{table}[h]
\caption{Critical point $g_{c}$. $g^{*}$ is the value of $g$ determined by
letting the stiffness constant $K_{\rho}={{1}\over{2}}$.}
\label{tab1}
\begin{center}
\begin{tabular}{lllllllll}
\hline\hline
$\ \ \ \ \ \ t/\omega_{0}$ &\ \  0.05 &\ \  0.1 &\ \  0.5 &
\ \  1 &\ \  5 &\ \  10 &\ \  20 & 100\\
\hline
$g_{c}/\omega_{0}$ (Ref.\cite{refa5}) & 2.297(2) & 2.093(2) & 1.63(1)
                              & 1.61(1) & 2.21(3) & 2.79(5) &  & \\
$g^{*}/\omega_{0}$ (Ref.\cite{refa5}) & 2.299 & 2.102 & 1.64 & 1.62
                              & 2.27 & 2.89 &  & \\
$g_{c}/\omega_{0}$ & 2.8215 & 2.1613 & 1.5939  & 1.6403  & 2.3068 & 2.8783
                         & 3.6868 & 6.9738 \\
\hline\hline
\end{tabular}
\end{center}
\end{table}

\begin{thebibliography}{45}
\bibitem{refa1}T. Holstein, Ann. Phys. {\bf 8}, 325 (1959).
\bibitem{refa2}R. H. McKenzie and J. W. Wilkins, Phys. Rev. Lett. {\bf 69},
               1085 (1992); K. Kim, R. H. McKenzie, and J. W. Wilkins,
               Phys. Rev. Lett. {\bf 71}, 4015 (1993).
\bibitem{refa3}E. Fradkin, and J. E. Hirsch, Phys. Rev. {\bf B27}, 1680
                 (1983).
\bibitem{refa9}T. W. Hagler and A. J. Heeger, Phys. Rev. {\bf B49}, 7313
                (1994).
\bibitem{refa4}H. Zheng, Phys. Rev. {\bf B56}, 14414 (1997).
\bibitem{refb1}J. E. Hirsch and E. Fradkin, Phys. Rev. {\bf B27}, 4302 (1983).
\bibitem{refb2}R. H. McKenzie, C. J. Hamer, and D. W. Murray, Phys. Rev.
                 {\bf B53}, 9676 (1996)
\bibitem{refb3}C. Bourbonnais and L. G. Caron, J. Physique {\bf 50}, 2751
                 (1989).
\bibitem{refb4}D. Schmeltzer, J. Phys. {\bf C18}, L1103 (1985).
\bibitem{refb5}H. Zheng, D. Feinberg, and M. Avignon, Phys. Rev. {\bf B39},
                9405 (1989).
\bibitem{refb7}G. Wellein and H. Fehske, Phys. Rev. {\bf B56}, 4513 (1997).
\bibitem{refa5}R. J. Bursill, R. H. McKenzie and C. J. Hamer,
              Phys. Rev. Lett. {\bf 80}, 5607 (1998).
\bibitem{refa6}A. Weisse and H. Fehske, Phys. Rev. {\bf B58}, 13526 (1998).
\bibitem{refa7}H. Zheng and M. Avignon, Phys. Rev. {\bf B58}, 3704 (1998).
\bibitem{refa13}L. G. Caron and C. Bourbonnais, Phys. Rev. {\bf B29}, 4230
                 (1984).
\bibitem{refa14}C. Zhang, E. Jeckelmann, and R. White, Phys. Rev. {\bf B60},
                 14092 (1999).
\bibitem{refa15}J. T. Gammel, I. Batistic, A. R. Bishop, {\it et al.},
                 Physica {\bf B163}, 458 (1990).

\end{thebibliography}
\end{document}